# Reducing US Biofuels Requirements Mitigates Short-term Impacts of Global Population and Income Growth on Agricultural Environmental Outcomes

David R. Johnson[a,b], Nathan B. Geldner[a], Jing Liu[c], Uris Lantz Baldos[c], Thomas Hertel[c]

[a] Purdue University, School of Industrial Engineering, West Lafayette, Indiana, USA
[b] Purdue University, Department of Political Science, West Lafayette, Indiana, USA
[c] Purdue University, Department of Agricultural Economics, West Lafayette, Indiana, USA

## Abstract

Biobased energy, particularly corn starch-based ethanol and other liquid renewable fuels, are a major element of federal and state energy policies in the United States. These policies are motivated by energy security and climate change mitigation objectives, but corn ethanol does not substantially reduce greenhouse gas emissions when compared to petroleum-based fuels. Corn production also imposes substantial negative externalities (e.g., nitrogen leaching, higher food prices, water scarcity, and indirect land use change). In this paper, we utilize a partial equilibrium model of corn-soy production and trade to analyze the potential of reduced US demand for corn as a biobased energy feedstock to mitigate increases in nitrogen leaching, crop production and land use associated with growing global populations and income from 2020 to 2050. We estimate that a 23% demand reduction would sustain land use and nitrogen leaching below 2020 levels through the year 2025, and a 41% reduction would do so through 2030. Outcomes are similar across major watersheds where corn and soy are intensively farmed.

## Introduction

Biofuels are an important component of United States energy and climate change mitigation policies. Forty-one states have adopted Renewable Portfolio Standards mandating minimal levels of electricity production from renewable sources for which at least one form of biomass qualifies (National Conference of State Legislatures, 2021). At the federal level, a Renewable Fuel Standard (RFS2) was first legislated in the Energy Policy Act of 2005 (US Federal Register, 2005) and expanded in the Energy Independence and Security Act of 2007 (US Federal Register, 2007). While these policies typically mention energy security and economic development as motivators, the primary stated motive is to reduce greenhouse gas (GHG) emissions from energy production. To this end, corn starch-based ethanol and other conventional biofuels are expected to comprise well over half of RFS2's proposed volume requirement of 20.77 billion gallons of ethanol-equivalents for 2022 (US Environmental Protection Agency, 2021).

However, analyses conclude that corn ethanol has comparable greenhouse gas emissions over the life cycle to petroleum-based fuels (Farrell et al., 2006). The fertilizer applied to increase corn yields exacerbates hypoxic "dead zones" in the ocean, chiefly the Gulf of Mexico (Diaz and Rosenberg, 2008; Rabalais and Turner, 2019), and leads to harmful algal blooms in the Great Lakes (Brooks et al., 2016; Smith et al., 2015; Watson et al., 2016). Liu, et al (2018) examine multiple interventions designed to meet the Hypoxia Task Force's target of a 45 percent reduction in nitrogen and phosphorus fluxes to the Gulf of Mexico, concluding that no single measure can effectively reach that goal (Liu et al., 2018; US EPA, 2016). Donner and Kucharik (2008) demonstrate that RFS2 mandates for liquid renewable fuel production make it "practically impossible [to meet hypoxia goals] without large shifts in food production and agricultural management."

Biobased energy feedstock production increases demand for agricultural land, a major driver of deforestation globally; RFS2 therefore is responsible for substantial net carbon emissions associated with indirect land use change (Busch and Ferretti-Gallon, 2020). As a result of the same dynamics, corn-based biofuel production may also lead to an increase in global food prices (Carter et al., 2017), thereby worsening global malnutrition, which remains a serious issue globally (Black et al., 2008; Gómez et al., 2013)[1]. The water required to grow corn for biofuel also imposes environmental costs, given serious water scarcity in some intensively cultivated regions of the United States (Shah et al., 2007). Finally, expansion of agricultural area and intensification of production on existing cropland leads to reduced biodiversity and ecosystem services, albeit in a complex fashion (Seppelt et al., 2016). To summarize, RFS2 results in production of corn ethanol which contributes relatively little to climate change mitigation objectives and imposes many other environmental costs.

All of these negative externalities are likely to be exacerbated by projected population and income growth, leading to greater demand for agricultural outputs (Riahi et al., 2017). In this paper, we examine the potential for reductions in the RFS2 mandate for conventional biofuels to mitigate the short-term environmental stressors of agricultural demand growth imposed by population and income growth. Our analysis utilizes a Gridded version of the Simplified International Model of agricultural Prices, Land use, and the Environment (SIMPLE-G) focusing on corn and soy production in the United States (SIMPLE-G-US-CS). SIMPLE-G is a partial equilibrium modeling framework for agricultural production and trade that can be used to evaluate agricultural policies while representing the heterogeneity of natural resources and productivity at high resolution (Baldos et al., 2020). Liu, et al. (2022) used the SIMPLE-G-US-CS model to examine the potential for various nitrogen management interventions to reduce leaching within the Mississippi River Basin, underscoring the need for high-resolution gridded analysis.

However, we stress that repeal of RFS2 mandates would not eliminate demand for corn ethanol. Babcock (2013) estimated an incremental 13.5% increase in demand for corn ethanol from RFS2, which Carter, et al. (2017) argues actually represents an upper bound on the decrease that would be associated with its repeal. Moschini, et al. (2016) instead estimates a demand reduction of 71% if RFS2 had been repealed in 2015; given consumption of 14.55 billion gallons of ethanol as a gasoline oxygenate in 2019 (i.e., pre-COVID-19), such a large reduction seems unlikely today. Currently, about 40% of corn grown in the US is used by ethanol production ("USDA ERS - Feedgrains Sector at a Glance," n.d.). The Biden administration's waiver in April 2022 allowing gasoline to be blended with up to 15% ethanol from June 1 to September 15 (when it is typically restricted in the US due to air quality concerns) (US EPA, 2022a), and subsequent increase in the ethanol volume required to be blended into gasoline in 2022 (US EPA, 2022b), illustrate the deep uncertainty in the actual demand reduction that would result from repeal of RFS2. In this paper, we therefore frame the results as an analysis of the outcomes associated with reduced US demand for corn starch-based ethanol, rather than the direct result of policy repeal.

## Methods

For this analysis, we employ a gridded version of the Simplified International Model of agricultural Prices, Land use, and the Environment (SIMPLE-G) (Baldos et al., 2020). This new version, SIMPLE-G-US-

---

[1] Not all studies attribute changes in food prices to the RFS2 policy specifically, such as Taheripour, et al. (2020) which concludes the long-term impact of RFS2 has been negligible.

CS, focuses on corn and soy production in the United States and is calibrated to a baseline year of 2010. It is a partial-equilibrium modeling framework for agricultural production and trade which can be used to evaluate agricultural policies while representing the spatial heterogeneity of natural resources and productivity at high resolution. Grid cells in the contiguous US are spaced at a 5 arc-minute resolution, with the rest of the world represented by 15 non-gridded regions (denoted in Figure 1). RFS2 biobased energy mandates, as well as demand in other regions, are modeled as exogenous demand shocks, consistent with the implementation of other parameters varied in our experimental design.

In SIMPLE-G-US-CS, baseline acreages of corn and soy production in each grid cell are derived from downscaling national data using the spatial pattern provided by the USDA Crop Data Layer data set. Separate acreages are specified for irrigated and rainfed production, and different transfer functions are used describing the relationship between nitrogen fertilizer application and yield for irrigated and rainfed areas. The yield transfer functions in Gompertz form are fitted to the pairs of nitrogen fertilizer application rate and the corresponding crop yield simulated by a process-based agro-ecosystem model named Agro-IBIS (Sacks and Kucharik, 2011). Similarly, leaching transfer functions in quadratic form are built into SIMPLE-G-US-CS to capture the nonlinear leaching response to various intensity of nitrogen fertilizer application. More details about the SIMPLE-G-US-CS model can be found in supplementary materials and Liu, et al. (2022).

Projections of population, income and total factor productivity growth, by region, correspond to Shared Socioeconomic Pathway 2 (SSP2), the IPCC's "Middle of the Road" pathway; population and income growth paths are shown in Figure 1. Biofuels demand projections in the rest of the world are aggregated regionally from the International Energy Agency's World Energy Outlook (International Energy Agency, n.d.). We modeled these changes in the years 2020 to 2050 at five-year intervals, combined with reductions in US biobased energy feedstock demands from 0% (i.e., no change) to 50% at 1-percent intervals. Total factor productivity changes for livestock, crops, and processed foods are taken from future projections from Ludena et al (2007), and historical estimates from USDA-ERS (2021) and Griffith et al (2004), respectively.

Gridded outputs from the SIMPLE-G-US-CS model were captured for land use (i.e., acreage of land under corn-soy production in each grid cell), corn-soy production, and associated nitrogen leaching. We also recorded the equilibrium corn-soy prices as calculated in each region.

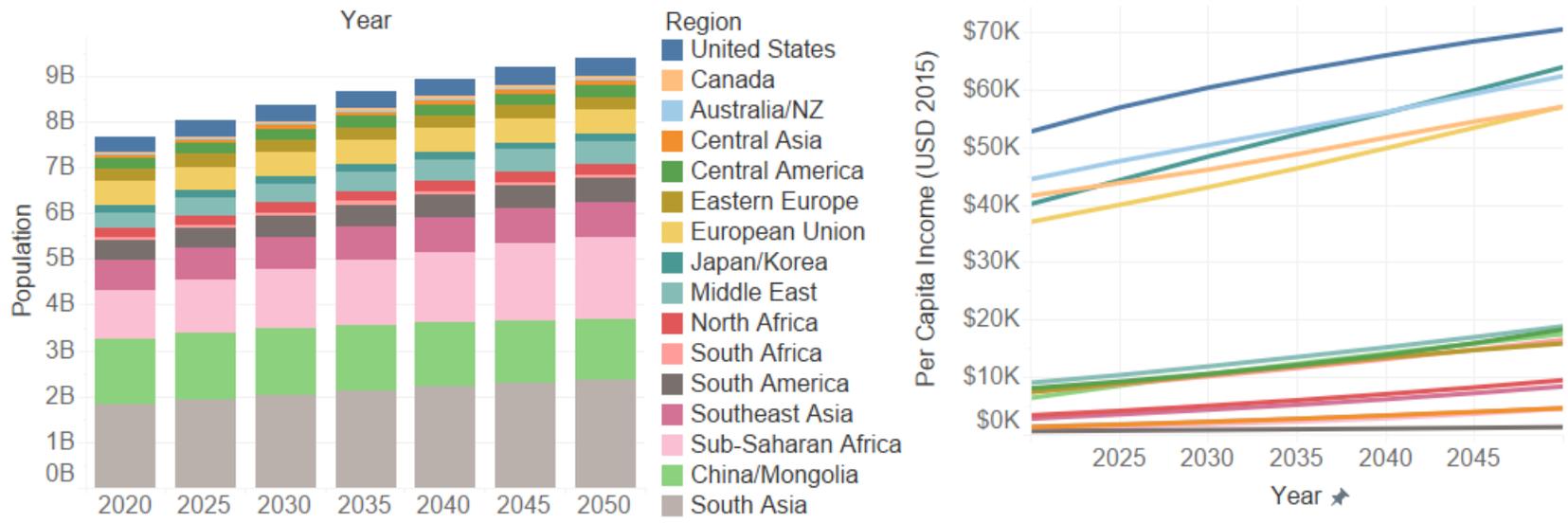

*Figure 1. Population over time (left) and per capita income (right), 2020-2050, SSP2, aggregated by SIMPLE-G-US-CS region*

## Results and Discussion

Aggregate US outcomes for land use, corn-soy production, and nitrogen leaching are presented in two different forms to highlight the role of RFS2-related demand reductions in future outcomes. Figure 2 presents the projections for growth in land use, production and leaching, relative to the 2020 benchmark. Figure 3 isolates the effect of the demand reductions by comparing future outcomes to a business-as-usual case, with 0% representing the *status quo* (i.e., no change to RFS2) and additional traces showing the deviations owing to RFS2 policy actions. For visual clarity, traces are only shown for demand reductions from 5 to 50 percent, at 5 percent intervals. We see total production and leaching outcomes that have similar sensitivity to biofuels demand reductions, with elasticities of approximately 0.4 and 0.5, respectively (i.e., a 10% reduction in the demand reduces production by 4% and leaching by about 5%). Land use is considerably less sensitive, with an elasticity of approximately 0.2.

The ability of the RFS2 repeal to mitigate the short-term impacts of population and income growth depends not only on the elasticity but on the current trends in environmental outcomes. In order to keep nitrogen leaching and land use below their 2020 levels by the year 2025, repeal would need to result in a demand reduction of at least 24%; under these demand reductions, overall corn-soy production in 2025 would still increase by 4.2%. To maintain nitrogen leaching and land use below their 2020 levels until the year 2030 would require a demand reduction of at least 41%; in this scenario, corn-soy production in 2025 would be 4.4% less than 2020 levels but 9% higher than 2020 levels in 2030.

When compared to outcomes in a *status quo* future, production, leaching, and land use all decline in response to a reduction in biofuels demand, with the proportional difference generally increasing over time. The difference grows more quickly in the first decade than from 2030-2050, although the distinction is slight; change in the difference over time is approximately linear.

The geospatial pattern of differences in outcomes is consistent across metrics and years, with selected results shown in Figure 4 and Figure 5. These figures depict nitrogen leaching outcomes in 2030, 2040, and 2050 associated with demand reductions of 24% and 41%, the reductions associated with maintaining leaching and land use outcomes below 2020 levels until 2025 and 2030, respectively. One interesting finding is that the areas with the greatest reduction in leaching when compared to the *status quo*, such as southwest Kansas and near the Appalachian Mountains, are the same areas with the largest leaching increases when expressed as a percentage of 2020 levels. These are marginal areas which have proven sensitive to demand fluctuations in the past (Lark et al., 2015).

Despite the limited spatial heterogeneity of outcomes in intensively farmed regions, it is nonetheless of interest to summarize the results by major basin, particularly nitrogen leaching which contributes to significant leachate exports to key water bodies (Ator et al., 2020; Ator and Denver, 2015; Donner and Kucharik, 2008; Mooney et al., 2020; Sun et al., 2020). Table 1 and Table 2 present the land use, corn-soy crop production, and nitrogen leaching over time, expressed both relative to the *status quo* and as a percentage of 2020 levels, respectively, in the Mississippi River Basin, Great Lakes Basin, and Chesapeake Bay. From these results, it is clear to see, for example, that even a large 41% reduction in US demand for biobased energy feedstocks will not by itself meet the Hypoxia Task Force's target 45% reduction in nitrogen fluxes to the Gulf of Mexico from the Mississippi River. Demand reductions have the greatest impact in the Chesapeake Bay region, followed by the Mississippi River and Great Lakes, although the differences across basins are not substantial.

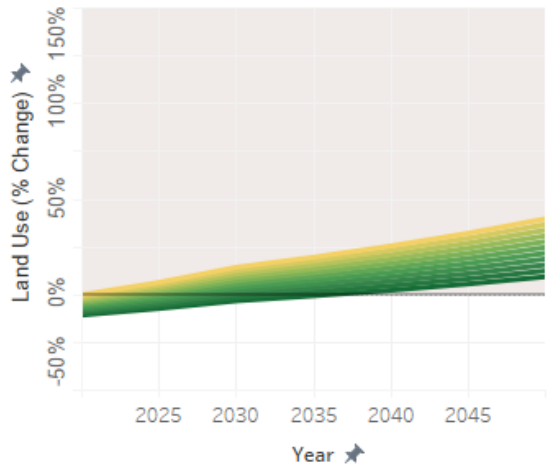 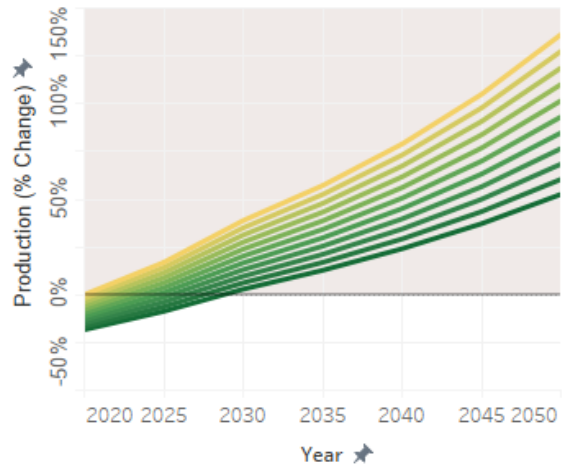 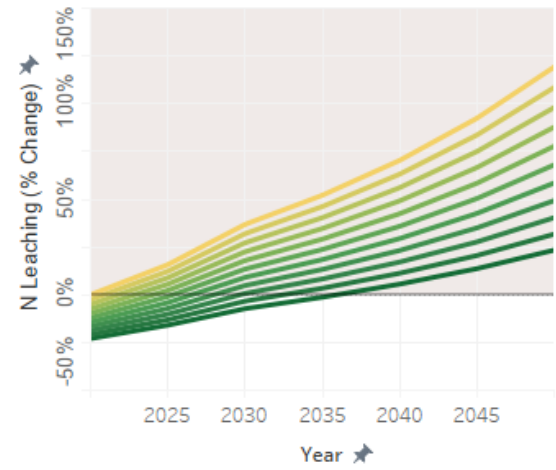

*Figure 2. Aggregated United States outcomes for corn-soy land use, production and leaching in the continental US, 2020-2050, relative to 2020 levels*

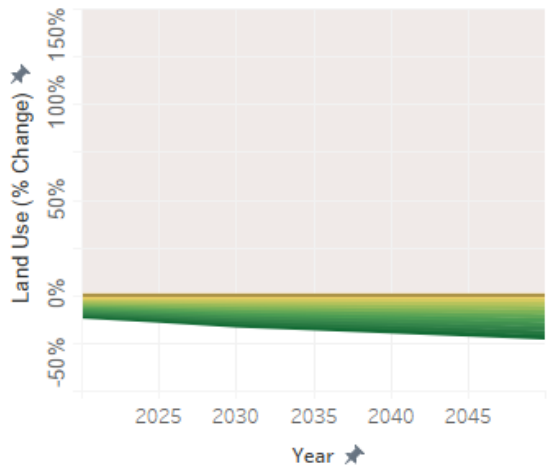 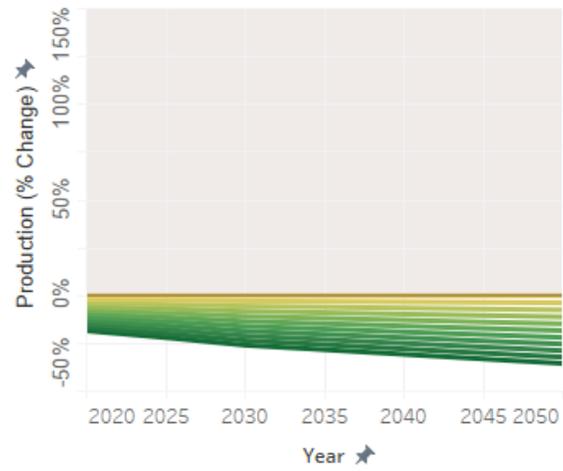 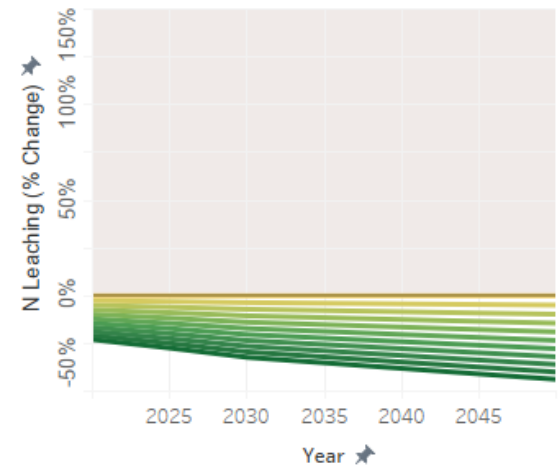

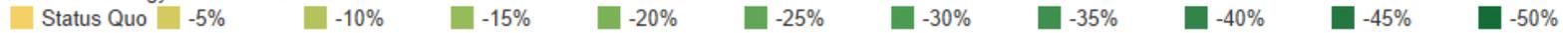

*Figure 3. Aggregated United States outcomes, 2020-2050, relative to status quo*

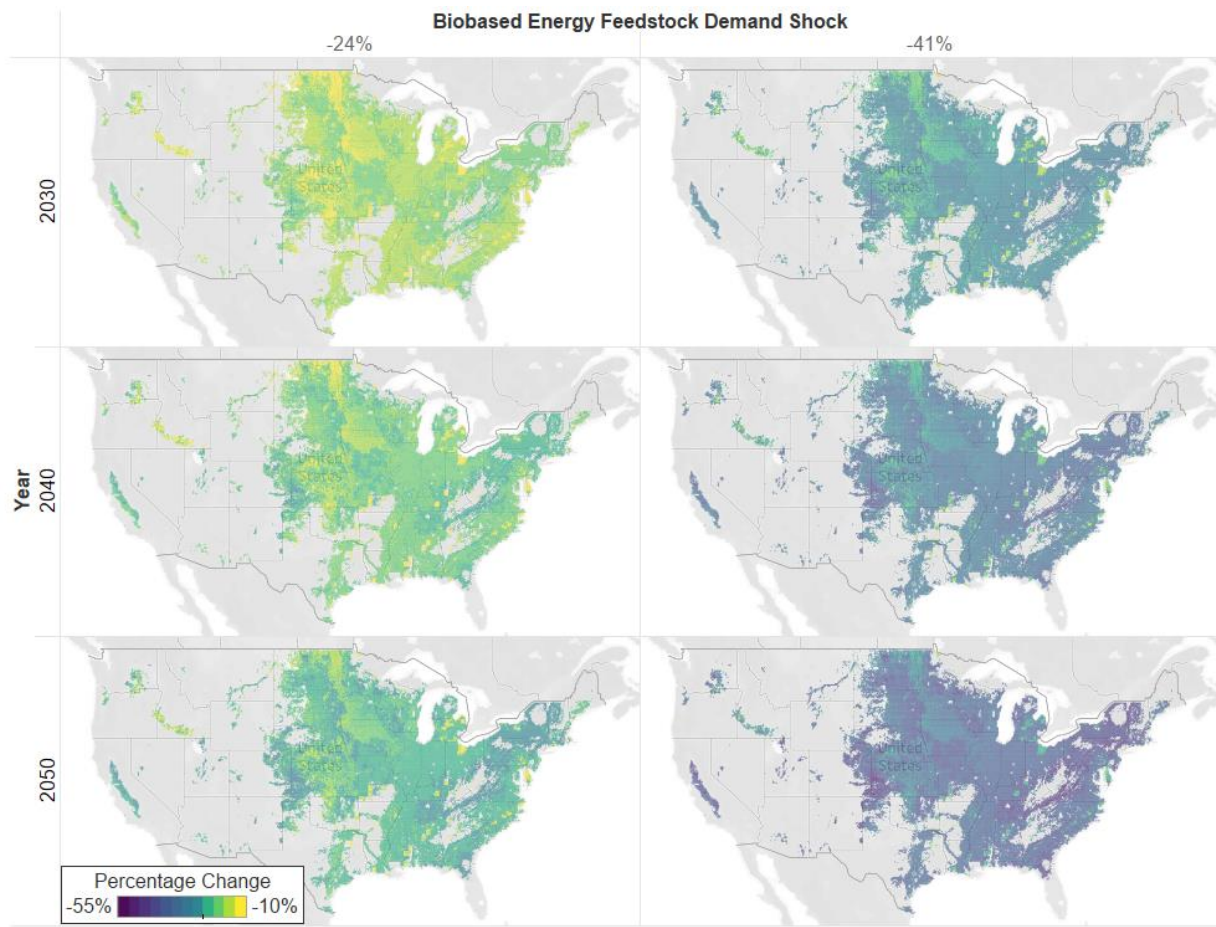

*Figure 4. Change in nitrogen leaching from corn-soy production with selected demand shocks, 2030 to 2050, relative to status quo*

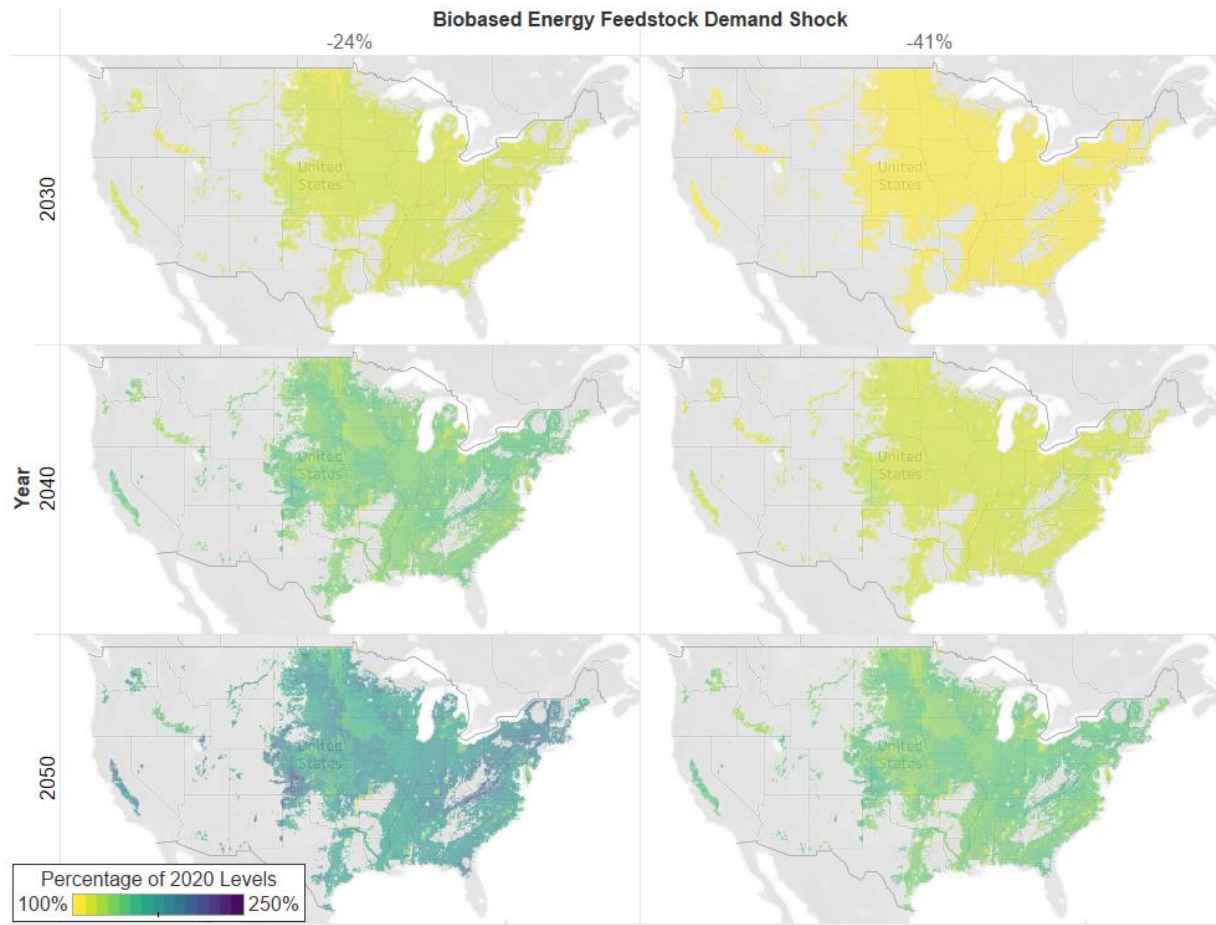

*Figure 5. Nitrogen leaching with selected demand shocks, 2030 to 2050, as a percentage of 2020 levels*

Table 1. Changes in outcomes in selected basins resulting from 24% and 41% demand reductions, relative to status quo

|  |  | Biobased Energy Feedstock Demand Shock / Year | | | | | |
|---|---|---|---|---|---|---|---|
|  |  | -24% | | | -41% | | |
|  |  | 2030 | 2040 | 2050 | 2030 | 2040 | 2050 |
| Mississippi River | Land Use | -7.5% | -8.9% | -10.3% | -12.8% | -15.3% | -17.9% |
|  | Crop Production | -12.9% | -15.2% | -17.5% | -21.5% | -25.5% | -29.4% |
|  | Nitrogen Leaching | -16.6% | -19.6% | -22.6% | -27.2% | -32.2% | -37.0% |
| Great Lakes | Land Use | -7.0% | -8.4% | -9.8% | -11.9% | -14.4% | -16.9% |
|  | Crop Production | -12.6% | -15.0% | -17.3% | -21.0% | -25.0% | -29.0% |
|  | Nitrogen Leaching | -15.8% | -18.9% | -21.9% | -26.0% | -30.9% | -35.9% |
| Chesapeake Bay | Land Use | -8.1% | -9.7% | -11.4% | -13.7% | -16.6% | -19.5% |
|  | Crop Production | -14.0% | -16.6% | -19.3% | -23.2% | -27.6% | -31.9% |
|  | Nitrogen Leaching | -16.7% | -20.1% | -23.5% | -27.4% | -32.7% | -38.1% |

*Table 2. Changes in outcomes in selected basins resulting from 24% and 41% demand reductions, as a percentage of 2020 levels*

| | | Biobased Energy Feedstock Demand Shock / Year | | | | | |
|---|---|---|---|---|---|---|---|
| | | -24% | | | -41% | | |
| | | 2030 | 2040 | 2050 | 2030 | 2040 | 2050 |
| Mississippi River | Land Use | 106.0% | 114.7% | 125.8% | 99.9% | 106.7% | 115.2% |
| | Crop Production | 121.0% | 151.7% | 194.9% | 109.0% | 133.4% | 166.8% |
| | Nitrogen Leaching | 114.5% | 137.5% | 170.7% | 99.9% | 116.1% | 139.0% |
| Great Lakes | Land Use | 105.6% | 113.6% | 123.9% | 99.9% | 106.1% | 114.1% |
| | Crop Production | 120.6% | 150.8% | 193.1% | 109.0% | 133.0% | 165.8% |
| | Nitrogen Leaching | 113.6% | 135.2% | 166.3% | 99.9% | 115.1% | 136.6% |
| Chesapeake Bay | Land Use | 106.4% | 115.9% | 128.2% | 99.9% | 107.1% | 116.5% |
| | Crop Production | 122.0% | 155.0% | 202.4% | 109.0% | 134.7% | 170.6% |
| | Nitrogen Leaching | 114.5% | 137.7% | 172.0% | 99.9% | 116.1% | 139.3% |

## Conclusions

Figure 2 illustrates the large increases from 2020 to 2050 in US corn-soy production, associated land use and nitrogen leaching projected under population and income changes corresponding to the SSP2 scenario. This poses a substantial challenge to the prospects of meeting environmental goals, such as reducing nitrogen fluxes to the Gulf of Mexico to manage the hypoxic zone. We have shown that repealing the US RFS2 mandates for liquid renewable fuel production would not achieve such a goal on its own, but demand reductions for US biobased energy feedstocks could mitigate the short-term impacts of population and income growth over the next decade. Overall corn-soy production would still rise over time in such a scenario thanks to the US Heartland's general productivity advantage compared to the rest of the world. This could make repeal of RFS2 a meaningful contributor to environmental goals when combined with other changes to management practices and policy interventions (e.g., carbon pricing, restoring depressional wetlands).

## Acknowledgments
This work was funded by the US National Science Foundation, award 1639268. The funding organization had no role in any part of this analysis.

## Supplementary Material: The SIMPLE-G-US-CS model

SIMPLE-G-US-CS is a global partial equilibrium economic model focusing on grid-specific crop productions and agricultural inputs use. It is a specialized version of the SIMPLE-G-US model (Baldos et al., 2020) focusing on just two crops: corn and soybeans. Because the prices for these two crops tend to move in tandem (they are close substitutes, both in production and in use), they can be combined into a single composite crop for purposes of market analysis. This corn-soy composite is obtained by converting soybean production into [price-weighted] corn-equivalent tons. This crop aggregation avoids the complexity arising from the cross elasticities of demand and supply over the long run (i.e., the relationship between the two products when the price of one of them changes), which is not the focus of the model. At each grid cell, crop production and input usage reflect different types of production (e.g., continuous corn, corn-soy rotation, etc.). The production functions are specific to each activity/grid cell, but all of them follow a multi-nesting input structure shown in Figure S1. Each nest represents a sub-production function with the constant elasticity of substitution functional form. The nesting structure differs at the bottom layer between irrigated and rainfed production. For irrigated crops, irrigation water is combined with irrigable land to produce a land-water composite, which is further combined with non-land inputs (e.g. capital and labor) to produce an augmented land input that is finally combined with N fertilizer to produce the ultimate crop output. The nitrogen fertilizer application rate data were provided by Lark, et al. (Lark et al., 2022).

The parameters labeled $\sigma$ are the elasticities of substitution between the two inputs within each layer. A larger value of $\sigma$ indicates easier substitution between inputs. When the use of one input is restricted by policy or natural constraints, the other input from the same layer will be employed more intensively, with the magnitude of this response governed by the size of the substitution elasticity and the extent of the change in relative input scarcity (relative prices). The production function—including nitrogen fertilizer intensity, input cost shares, and substitutability of inputs—is unique for each grid cell.

Given the focus on nitrogen fertilizer (N) in this work, we pay special attention to the relationship between N and crop production, as well as that between N and leaching. For this purpose, we build on the work of Liu, et al. (2018) which fit transfer functions to the outputs of the Agro-IBIS agro-ecosystem model (Kucharik, 2003). Figure S2 shows how this is done for one practice in a single grid cell. By running Agro-IBIS many times (individual dots), we are able to fit a so-called Gompertz curve to the relationship between N applications and crop yield, as well as a quadratic leaching curve characterizing the relationship between N applications under a given practice in a given location and leaching out of the root zone. These transfer functions are incorporated into SIMPLE-G-US-CS following the approach of Liu, et al. (2018), allowing for a spatially resolved characterization of the tradeoffs between agricultural production and the environment in the SIMPLE-G-US-CS model.

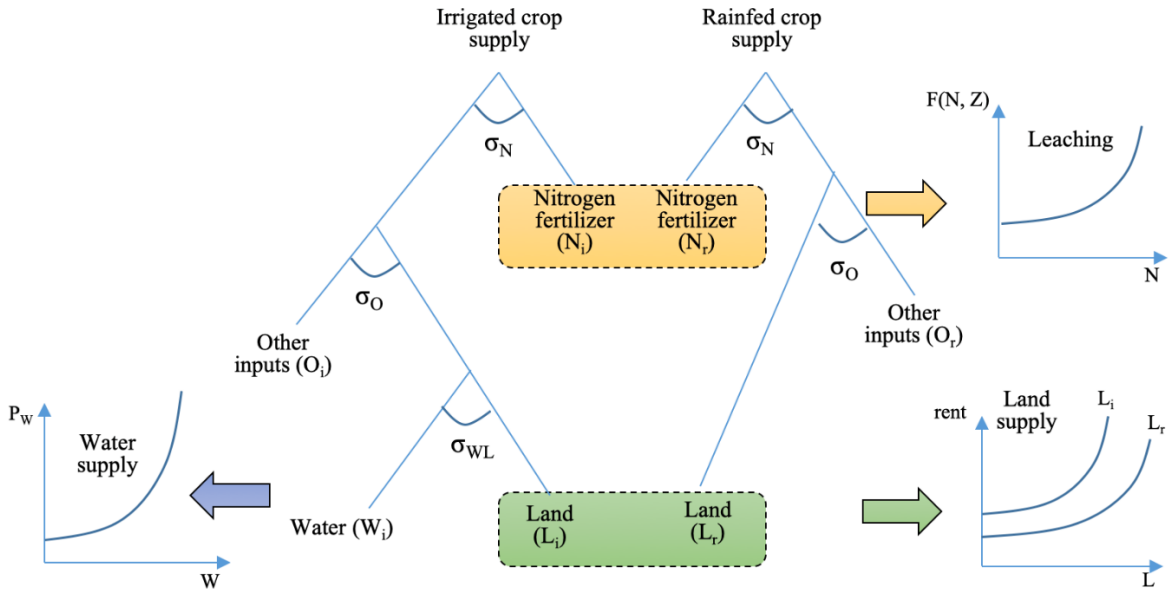

*Figure S1. Conceptual diagram illustrating the output structure of the SIMPLE-G-US-CS model*

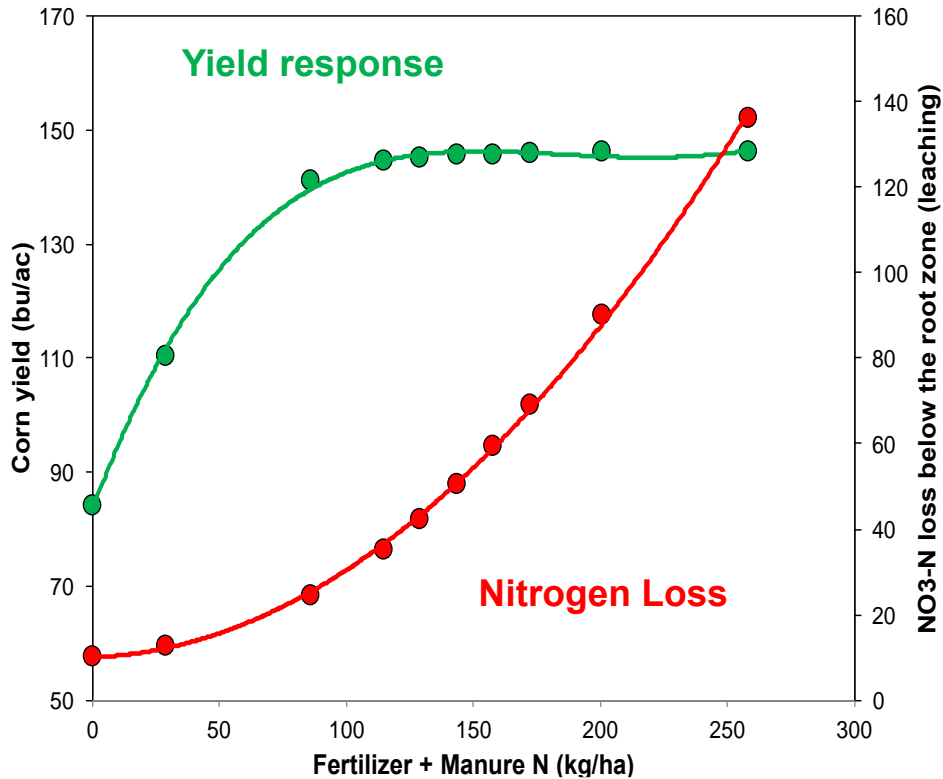

*Figure S2. Yield response (Gompertz function) and leaching response (quadratic function) to N fertilizer application rate developed based on Agro-IBIS simulations.*

The demand for the corn-soy composite is driven by biofuels production, and the consumption of crops, processed foods and livestock products, which is in turn driven by population and per capita income.

The equilibrium for sluggish and semi-mobile inputs (e.g., cropland and water) is established at the grid-cell level, while the markets for mobile inputs (e.g., labor and capital) are cleared at the regional level. More descriptions of the SIMPLE-G-US model can be found in Baldos, et al. (2020) and Sun, et al. (2020).

The baseline scenario to 2050 builds on the business as usual shared socioeconomic pathway (SSP2) (O'Neill et al., 2014; Riahi et al., 2017). In addition to the growth rates for population and income from SSP2, we incorporate biofuels growth rates estimated by the IEA (International Energy Agency, n.d.) and agricultural productivity growth rates from Ludena, et al. (2007), with historical estimates from USDA-ERS (2021) and Griffith, et al. (2004).